\documentclass[sigconf]{acmart}

\usepackage{booktabs} 
\usepackage{wrapfig}
\setcopyright{rightsretained}
\usepackage{booktabs}
\usepackage{multirow}
\usepackage[colorinlistoftodos]{todonotes}
\usepackage[caption=false]{subfig}
\usepackage{multirow}

\usepackage{pifont}

\newcommand{\cmark}{\ding{51}}%
\newcommand{\xmark}{\ding{55}}%

\setcopyright{none}
\renewcommand\footnotetextcopyrightpermission[1]{} 
\acmArticle{4}
\acmPrice{15.00}

\PassOptionsToPackage{bookmarks=false}{hyperref}
\PassOptionsToPackage{draft}{hyperref}
\hypersetup{draft}

\begin{document}
\title{Performance Exploration of Virtualization Systems}

\author{Joel Mandebi Mbongue}
\affiliation{%
  \institution{University of Florida}
  \city{Gainesville}
  \country{Florida}}
 \email{jmandebimbongue@ufl.edu}
 
 \author{Danielle Tchuinkou Kwadjo}
\affiliation{%
  \institution{University of Florida}
  \city{Gainesville}
  \country{Florida}}
 \email{dtchuinkoukwadjo@ufl.edu}
 
 \author{Christophe Bobda}
\affiliation{%
  \institution{University of Florida}
  \city{Gainesville}
  \country{Florida}}
 \email{cbobda@ece.ufl.edu}


\begin{abstract}
Virtualization has gained astonishing popularity in recent decades. It is applied in several application domains, including mainframes, personal computers, data centers, and embedded systems. While the benefits of virtualization are no longer to be demonstrated, it often comes at the price of performance degradation compared to native execution. In this work, we conduct a comparative study on the performance outcome of VMWare, KVM, and Docker against compute-intensive, IO-intensive, and system benchmarks. The experiments reveal that containers are the way-to-go for the fast execution of applications. It also shows that VMWare and KVM perform similarly on most of the benchmarks.
\end{abstract}

\maketitle
\section{Introduction}

Virtual machines (VM) have been introduced early in the 1960s by IBM to consolidate the hardware and decrease exploitation costs \cite{denning2001anecdotes}. The mainframes were sold at about \$2.9 million (equivalent to about \$25 million in 2020) and rented for \$63,500 (about \$553,417 in 2020) per month in a typical configuration, making computing systems only accessible to a small range of customers \cite{InflationCalculator, IBM_History}. A VM could be seen as an instance of the physical machine in which the users had the illusion of fully owning the hardware. In reality, it was just a way to transparently share resources and run workloads from different users in an isolated way on the same hardware. A few decades later, researchers investigated models, challenges, and solutions to efficiently implement “virtual sub-environments” in physical machines \cite{chiueh2005survey}. The VM abstraction then provided concurrent and interactive access to the underlying hardware.

The continuous innovation in virtualization technology has led to the emergence of an ecosystem of products ranging from VMs running on personal computers to enterprise and commercial systems running in the cloud. Virtualization concepts are also applied beyond traditional hardware devices such as processors, memory, disk, and network cards. As example, some research propose to virtualize Field-Programmable Gate Arrays (FPGA) for cloud and data center applications \cite{mbongue2018fpgavirt, mbongue2018fpga, mbongue2020architecture}. Graphic Processing Units (GPU) are also provisioned as part of virtual resource pools \cite{hong2017fairgv,hong2017gpu}. Among the most common virtualization softwares are VirtualBox, KVM, QEMU, Xen, VMware workstation, and container engines such as Docker and LXD.
The emergence of multiple virtualization systems supporting hardware consolidation in personal computers, embedded systems, and cloud-scale deployments raise the need for architecture classification and performance evaluation. In the context of this work, we study the architectures of state-of-the-art virtualization systems and provide a quantitative evaluation of the performance that can be achieved against IO-intensive (such as applications intensively accessing the disk), memory-intensive (such as matrix-based applications), and compute-intensive benchmarks (such as high-performance applications). We also evaluate the overhead introduced by virtualization technologies against native executions.

\begin{figure}[]
      \centering
      \includegraphics[width=.94\linewidth]{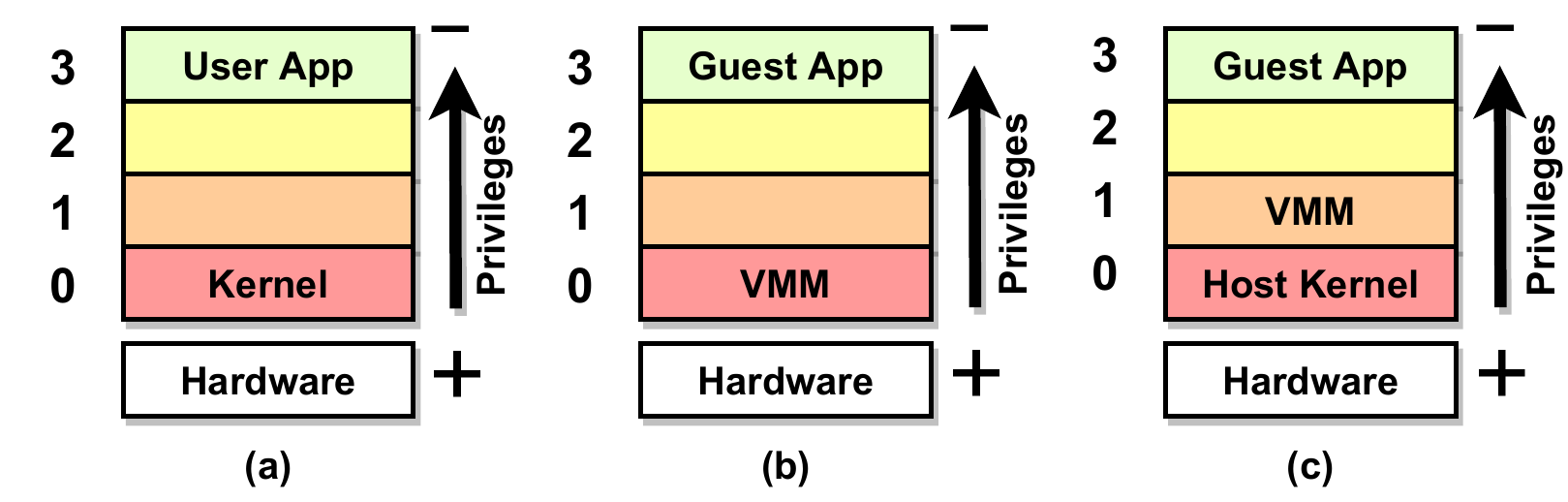}
      \caption{x86 Privilege Ring and Virtualization. (a) Typical configuration in environment with no virtualization. The kernel runs at level 0 and applications run at level 3. (b) Corresponds to bare-metal virtualization stacks. There is no host operating system, the virtual machine monitor (VMM) runs at level 0 and guest applications are at level 3. (c) Deployment of hosted VMMs. The host kernel runs at level 0, the VMM at level 1, and the guests at level 3.}
      \label{fig:x68_ring}
\end{figure}
\section{BACKGROUND}
\subsection{Type of Virtual Machine Monitors}
VMs have several advantages among which easy maintenance, fast recovery from fault, rapid  provisioning and domain isolation \cite{bugnion2017hardware}. They allow running multiple operating systems simultaneously on the same machine. Furthermore, they support the execution of systems with entirely different instruction set architectures than that of the underlying hardware. VMs typically run above a software called "\textit{Virtual Machine Monitor}" (VMM) or simply hypervisor. It controls the run-time resources of the VMs and ensures proper execution of privileged instructions.

The x86 architecture separates processor privileges with a protection ring or levels \cite{chirammal2016mastering}. It is a mechanism that protects data and restricts operations that programs can run. Each program that executes in an x86 system is assigned to a specific ring or level that defines the access privileges on system resources. Figure \ref{fig:x68_ring} shows the different privilege levels available in x86 architectures. Typically, level 0 is reversed for the operating system (OS) services that directly interface with the hardware (kernel mode). Levels 1 and 2 are mostly unused and are reserved for some drivers and middleware. User applications run at level 3 (user mode) \cite{chirammal2016mastering}. In Figure \ref{fig:x68_ring}(a), no virtualization is implemented. The user applications run at level 3 and the kernel of the OS handles privileged instructions at level 0. Executing at level 0 allow the kernel  to directly access and control the hardware. Depending on how far apart the VMM is from the actual hardware in the x86 privilege levels, we consider two types of hypervisors \cite{desai2013hypervisor, force2000analysis}: \textbf{(1) Type-1 hypervisors} (bare metal): the VMM is installed directly above the hardware (see Figure \ref{fig:x68_ring}(b)). Examples of such VMMs include Xen and Linux enabled by Kernel-based (KVM) \cite{KVM_redhat}. The VMM is responsible from emulating the privileged instructions launched in the guest space. \textbf{(2) Type-2 hypervisor (hosted)}: in this configuration, the VMM is installed in the host OS (see Figure \ref{fig:x68_ring}(c)). An example of this category is VMware Workstation. Privileged instructions in the guest space typically cause a "\textit{world switch}" to the host kernel under the supervision of the VMM. In general, a set of applications or/and drivers implemented in the VMM are used to access kernel privileged instructions.

\subsection{VMware Workstation}
\begin{figure}[]
     \centering
      \includegraphics[width=.6\linewidth]{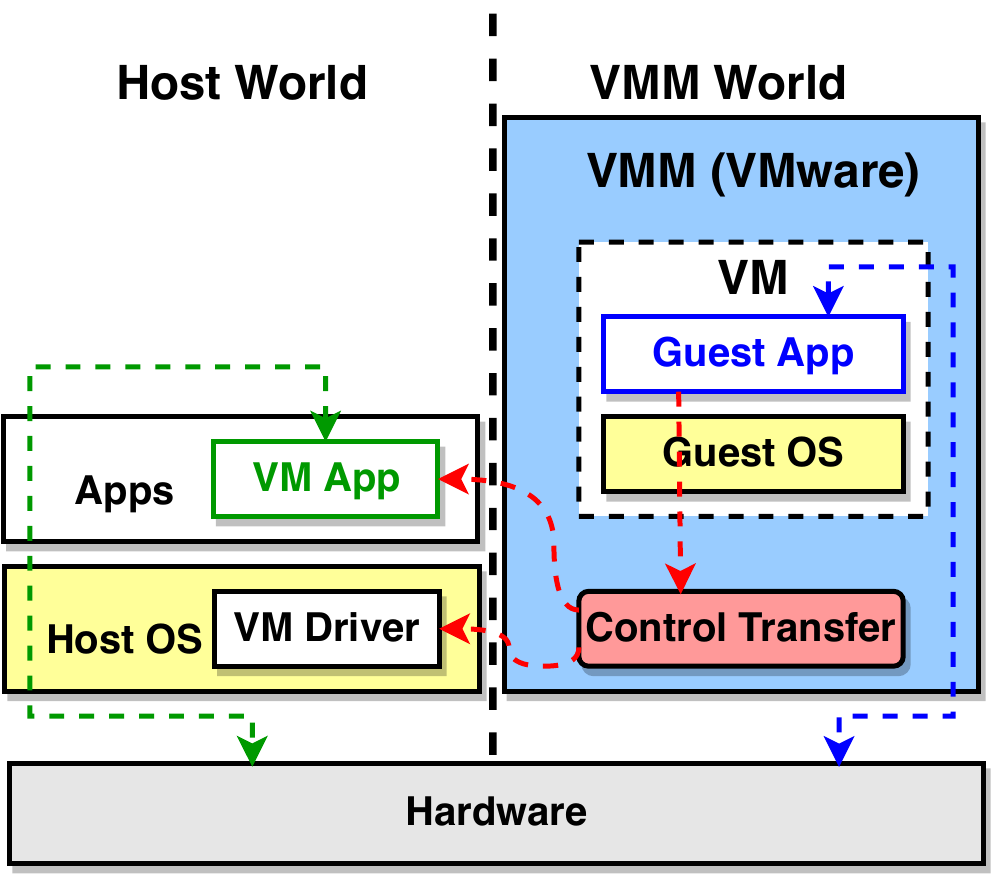}
      \caption{VMware Workstation Architecture}
      \label{fig:vmWare}
\end{figure}

VMware Workstation is a Type-2 hypervisor that runs on x86 processors. It supports Windows and Linux hosts, and allows users to run multiple VMs on a single machine \cite{bugnion2012bringing}.  It virtualizes IO devices using a hosted IO model which consists in taking advantage of pre-existing support in the host OS. This approach has several advantages among which application portability and consistency. It also delivers near native performance for CPU-intensive workloads. Figure \ref{fig:vmWare} summarizes the architecture of VMware workstation. Non-privileged instructions from the guest can run natively on the hardware without interference from the VMM. On the other hand, when guest applications issue privileged instructions, the VMM traps and emulates. Specifically, the VMM requests a "world switch" from the \textit{VM Driver}. Next, the VMM provides data to the \textit{VM App}.
The \textit{VM App} is then in charge of mapping the virtual requests to host system calls \cite{vmware_architecture}.
After completing the system calls, the \textit{VM Driver} returns the control to the VMM. The VMM collects the results from the \textit{VM App} and passes them to the VM. The VM can then resume its normal execution.

\subsection{Kernel-based Virtual Machine}
Kernel-based Virtual Machine (KVM) is a virtualization module present in Linux releases since kernel version 2.6.20. It represents the latest generation of open source virtualization utility. It transforms a Linux system into a Type-1 hypervisor and benefits from decades of innovation in Linux process scheduling, memory management, device drivers, etc --- to manage VMs \cite{KVM_redhat}. It requires processors with virtualization extensions such as Intel VT or AMD-V. To emulate processors and IO devices, KVM is combined with QEMU (Quick Emulator) \cite{chirammal2016mastering}. IO communication between the virtual and physical system is done through VirtIO. VirtIO is an abstraction of IO devices implemented by Rusty Russel for communication interfaces between guests and host in paravirtualized architectures. KVM uses VirtIO as paravirtualized device drivers since kernel version 2.6.25 \cite{russell2008virtio, chirammal2016mastering}.
\begin{figure}[]
     \centering
      \includegraphics[width=.6\linewidth]{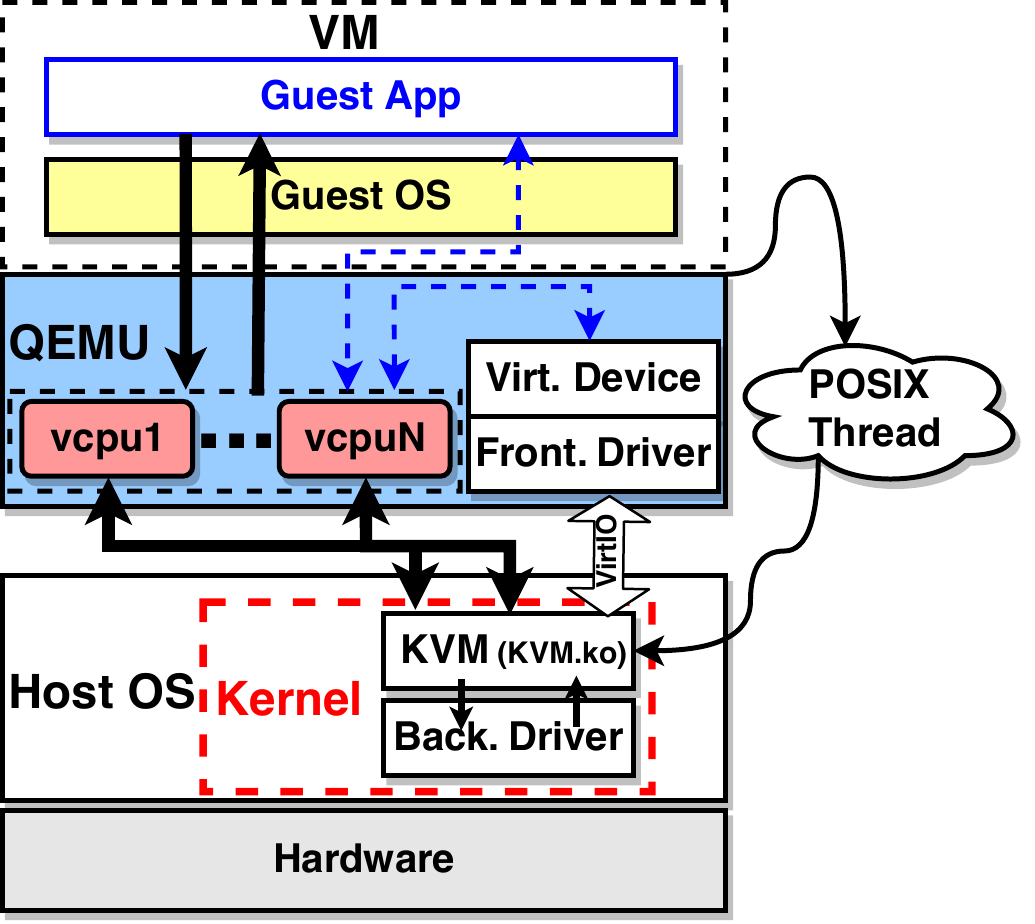}
      \caption{Overview of the KVM-QEMU Virtualization Architecture}
      \label{fig:qemuKVM}
\end{figure}
Figure \ref{fig:qemuKVM} highlights the key components of the KVM-QEMU virtualization. To execute guest applications on the physical hardware, QEMU creates POSIX threads that represent the virtual CPUs. It has the advantage of making virtual applications appear as processes in the host environment. The guest applications are run via KVM kernel modules that provide extension support for hardware virtualization such as Intel VMX \cite{uhlig2005intel}.
Specifically, QEMU opens the device file \textit{/dev/kvm} exposed by KVM kernel module and runs a set of \textit{ioctls()} functions. These functions allow setting and updating the state of the registers of each virtual CPU in the QEMU internal data structure, thus ensuring a smooth execution of guest applications \cite{chirammal2016mastering}. This whole emulation however comes with a considerable overhead. In a comparative study, Weber et .al reported that QEMU was up to 5$\times$ slower than native environment on some compute-intensive applications \cite{weber2013scientific}.

\begin{table*}[]
\caption{List of the testing applications}
\label{tab:my-benchmarks}
\begin{tabular}{|c|c|c|c|}
\hline
\textbf{Category} & \textbf{Benchmark \cite{Phoronix_test_suite}} & \textbf{Details}& \begin{tabular}[c]{@{}c@{}}\textbf{Scales with}\\ \# \textbf{CPU cores}\end{tabular}                                           \\ \hline\hline

\multirow{2}{*}{\textbf{Processor}} & \textit{pts/aobench}      & \begin{tabular}[l]{p{0.6\textwidth}}It returns the average execution time in seconds. The lower the value is, the better it is.\end{tabular} & \xmark                                                                                                                                                                                                                                                          \\ \cline{2-4} 
                                    & \textit{pts/asmfish}      & \begin{tabular}[l]{p{0.6\textwidth}}It stresses the processors. It returns a score that represents the average number of nodes per second. The higher the value is, the better it is.\end{tabular} &\cmark                                                                                                                                                          \\ \hline
\multirow{4}{*}{\textbf{Disk}}      & \textit{pts/aio-stress}   & \begin{tabular}[l]{p{0.6\textwidth}}Performs asyncrhonous IO operations on the disk.It returns the average the throughput in MB/s. The higher the value is, the better it is.\end{tabular} & \textbf{n/a}                                                                                                                                                                  \\ \cline{2-4} 
                                    & \textit{pts/blogbench}    & \begin{tabular}[l]{p{0.6\textwidth}}Mimics the load of real-world busy servers on the filesystem with random read, write,rewrite. It returns a score. The higher the value is, the better it is. The benchmark runs a write and a read test.\end{tabular} & \textbf{n/a}                                                                                                \\ \cline{2-4} 
                                    & \textit{pts/compilebench} & \begin{tabular}[l]{p{0.6\textwidth}}Simulate common IO operations on disk. It measures how well the filesystem can maintain directory locality as the disk fills up. It returns the throughput in MB/s. The higher the value is, the better it is. The benchmark  runs 3 different tests: Compile, Initial Create, and Read Compile Tree.\end{tabular} & \textbf{n/a}                \\ \cline{2-4} 
                                    & \textit{system/iozone}    & \begin{tabular}[l]{p{0.6\textwidth}}Tests the hard drive and file-system performance. It returns the average throughput in MB/s. The higher the value is the better it is. The benchmark assesses read and write performance. In our experiments, we tested with the following parameters: record size = 1MB and size = 512 MB.\end{tabular} & \textbf{n/a}            \\ \hline
\multirow{2}{*}{\textbf{System}}    & \textit{pts/phpbench}     & \begin{tabular}[l]{p{0.6\textwidth}}Performs a large number of simple tests in order to bench various aspects of the PHP interpreter. It returns a score. The higher the value is, the better is it.\end{tabular} & \xmark                                                                                                                                            \\ \cline{2-4} 
                                    & \textit{pts/stress-ng}    & \begin{tabular}[l]{p{0.6\textwidth}}It stresses a computing system in various ways. It returns a score in Bogo Ops/s that reflects how well the system reacted.  The higher the value is, the better it is. We ran 5 stressors that are: Memory Copy, Matrix Math, Vector Math, Context Switching, and Crypto\end{tabular} & \cmark                                  \\ \hline
\end{tabular}
\end{table*}

\subsection{Containers: Docker}
\subsubsection{Containers}
Containers are virtualization technologies in which the virtual environment directly runs above the host OS. They run within a container engine instead of an hypervisor. They are not designed to run a complete systems, but focus at the application level.
\begin{figure}[h]
     \centering
      \includegraphics[width=.8\linewidth]{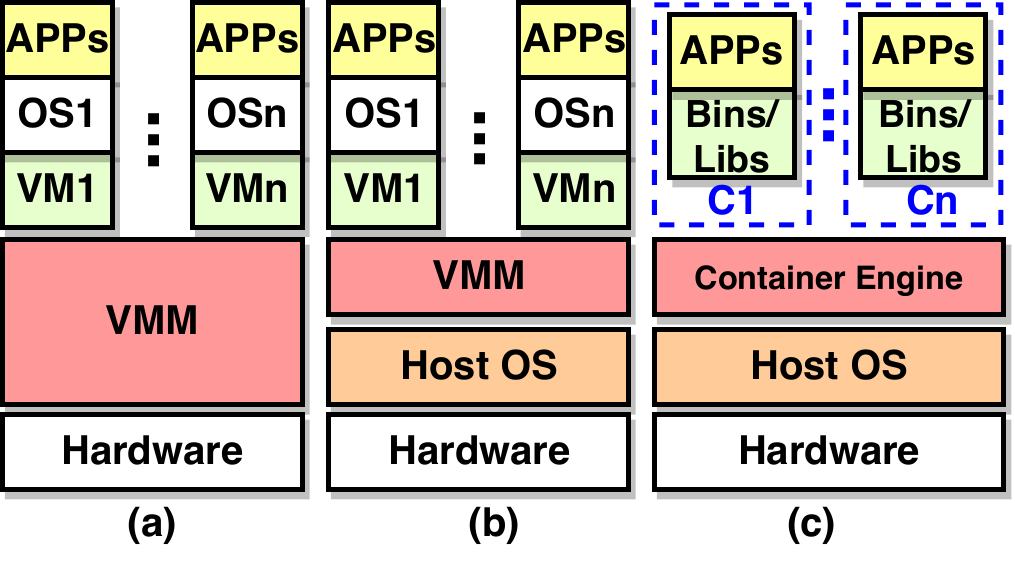}
      \caption{Difference between containers and VMs. (a) VMs running on a bare-metal hypervisor. (b) VMs deployment on a hosted hypervisor. (c) Deployment architecture of containers.}
      \label{fig:virt_containers}
\end{figure}
Containers are developed to reduce the footprint of systems, especially those that do not need heavy virtualization infrastructures. Figure \ref{fig:virt_containers}(a) and (b) show the typical virtualization stacks for VMs. Next, Figure \ref{fig:virt_containers}(c) illustrate the key difference between container and VM stacks. It resides in that containers only run applications on top of a \textit{container engine} instead of a hypervisor. Containers only need application binaries and a run-time engine, while VMs require support to run entire guest OSes above the underlying OS or hypervisor. They implement "\textit{OS-level virtualization}" as opposed to "\textit{hardware virtualization}" with VMs. This particular feature makes them lightweight and very portable. Containers are well-suited for fast development and deployment of applications as codes and dependencies can be packed and easily made available to users. They are nevertheless less flexible than VMs as they cannot run an entire OS. The \textit{container engine} runs as privilege level 3 in the x86 hierarchy, which means that all the containers in a machine share the same host kernel \cite{de2017survey}. However, this can cause multiple vulnerabilities that adversaries could exploit to bridge into the system.

\subsubsection{Docker}
Docker is one of the most popular container technology currently in use \cite{combe2016docker}. It mainly focuses on improving developer experience and enable the distribution of microservices as images for direct deployment.

 In the next section, we will discuss our methodology for evaluating the performance of the different virtualization technologies that we study.
\section{Method and approach}
Our approach to carry out the comparative study can be summarized in 4 steps:
  \begin{enumerate}
      \item \textbf{Categorizing the major virtualization schemes:} this first step was accomplished in the previous section. The purpose is to limit the scope of our study to a well-defined set of tools implementing the selected virtualization architectures.
      
      \item \textbf{Selecting the tools:}  in this work, we compare the performance of KVM, VMware workstation, and Docker as they represent examples of Type-1 and Type-2 hypervisors, and container.
      
      \item \textbf{Selecting the properties to assess:} we focus on evaluating how the different selected virtualization environments perform against some workloads. We will particularly observe IO speed by measuring how the disk access time scales with different applications. We will also study the memory consumption when running the same applications across the different environments. Finally, we will look at the processor utilization.
      
      \item \textbf{Recording and Analysing results:} we record observations from running the experiments. Next, we present the results and discuss the observed metrics.
      
  \end{enumerate}

\section{Experimental Evaluation}
In this section, we present and elaborate on the experimental observations.

\subsection{Evaluation Setup}
In order to conduct our experiments, we installed the 3 virtualization software stacks in a Dell R7415l EMC server running on a 2.09GHz AMD Epyc 7251 CPU $\times$16 cores with 64GB of memory and 1TB of hard drive. We installed CentOS-7 64-bit with a kernel of version 3.10.0 to manage the resources of the server. To run virtual machines on KVM, we installed Virtual Machine Manager 1.5.0 and QEMU 2.11.50. We also installed VWware Workstation 15.5.2 or VMWare for brevity. Next, we created a virtual machine with 8GB of RAM, 4 processors, and 40 GB of hard drive (SCSI). We installed the same release of CentOS-7 that was installed on the server. Finally, to conduct experiments on containers, we installed Docker version 1.13.1 on the Dell EMC server.

\subsection{Evaluation Applications and Platforms}

\subsubsection{Benchmark Suite:}
After the selection of the virtualization tools, one of the most critical task consists in selecting the set of testing applications. Because our first concern was to find applications that can run on all of our virtualization tools, we selected the Phoronix Test Suite v9.6.0 (Nittedal) \cite{Phoronix_test_suite}. We downloaded the stable release for Linux and pulled the corresponding Docker image from Docker Hub. It features more than 200 individual test profiles and more than 60 test suites. It provides an interactive command line interface (CLI) that allows running testing applications with well-defined attributes. Results can be saved under multiple formats such as HTML, PDF, and plain text. 
Table \ref{tab:my-benchmarks} provides the list of benchmarks we use in our experiments. We specifically check IO, processor, and system performance.

\subsubsection{Testing platforms:}
In all the experiments, we evaluate each of the benchmarks listed in Table \ref{tab:my-benchmarks} on four environments that are: native system, Docker, VMWare, and KVM. It evaluates the performance of the different virtualization systems and provides an overview of the virtualization overhead compared to applications running without virtualization.

\subsection{Experiment Analysis}
\subsubsection{Disk Access Performance}
\begin{figure}[h]
     \centering
      \includegraphics[width=.6\linewidth]{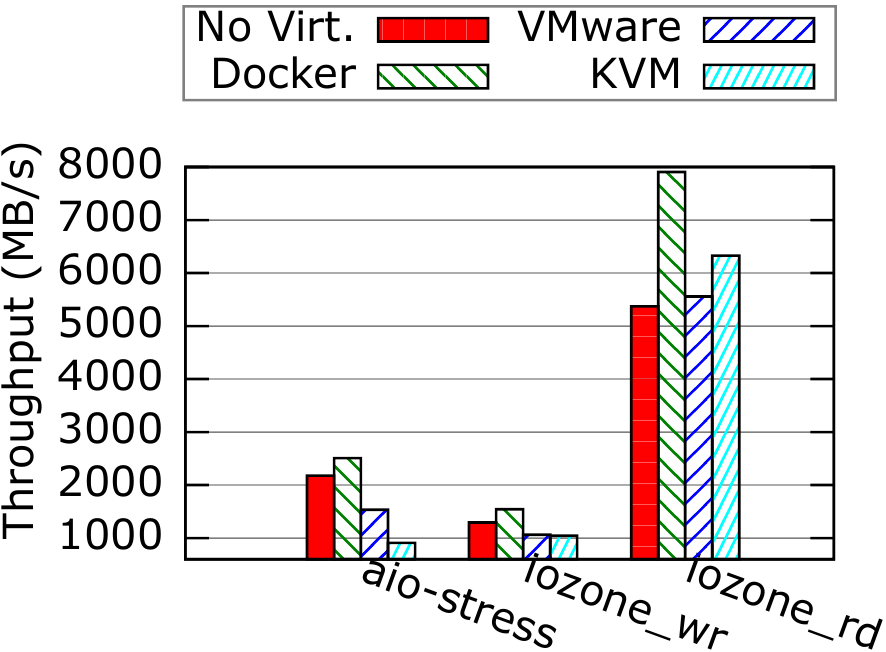}
      \caption{Combined disk performance metrics. "aio-stress" refers to the aio-stress test. "iozone\_wr" indicates the write performance test of the iozone benchmark. "iozone\_rd" specifies the read performance test of the iozone benchmark }
      \label{fig:diskperformance1}
\end{figure}

In this section, we discuss experiments related to comparing disk access performance.
Our first test focuses on stressing the file system using the \textbf{pts/aio-stress} and \textbf{system/iozone} benchmarks. The results from the executions are summarized in Figure \ref{fig:diskperformance1}.
We observed that in all of the three test applications, Docker achieved the highest throughput. Next comes the native execution, followed by VMware and KVM. One explanation is that containers do not need to "trap and emulate" as they run directly above the host OS, giving them a clear advantage over the virtualization with a hypervisor. Containers also outperformed the native execution consistently in the three experiments. This may result from the resource isolation implemented by containers, limiting interference from other processes in the system.

\begin{table*}[h]
\scriptsize
\caption{Stress-ng execution results}
\label{tab:stress_ng}
\begin{tabular}{|c|c|c|c|c|c|c|c|c|c|c|c|c|c|c|c|}
\hline
 & \multicolumn{3}{c|}{\textbf{Memory Copy}} & \multicolumn{3}{c|}{\textbf{Matrix Math}} & \multicolumn{3}{c|}{\textbf{Vector Math}} & \multicolumn{3}{c|}{\textbf{Context Switching}} & \multicolumn{3}{c|}{\textbf{Crypto}} \\ \cline{2-16} 
\multirow{-2}{*}{} & \textbf{Value} & \textbf{\begin{tabular}[c]{@{}c@{}}Memory \\ Usage\\ (MB)\end{tabular}} & \textbf{\begin{tabular}[c]{@{}c@{}}CPU\\ Usage\end{tabular}} & \multicolumn{1}{c|}{\textbf{Value}} & \multicolumn{1}{c|}{\textbf{\begin{tabular}[c]{@{}c@{}}Memory\\ Usage\\ (MB)\end{tabular}}} & \multicolumn{1}{c|}{\textbf{\begin{tabular}[c]{@{}c@{}}CPU\\ Usage\end{tabular}}} & \textbf{Value} & \textbf{\begin{tabular}[c]{@{}c@{}}Memory\\ Usage\\ (MB)\end{tabular}} & \textbf{\begin{tabular}[c]{@{}c@{}}CPU\\ Usage\end{tabular}} & \multicolumn{1}{c|}{\textbf{Value}} & \multicolumn{1}{c|}{\textbf{\begin{tabular}[c]{@{}c@{}}Memory\\ Usage\\ (MB)\end{tabular}}} & \multicolumn{1}{c|}{\textbf{\begin{tabular}[c]{@{}c@{}}CPU\\ Usage\end{tabular}}} & \multicolumn{1}{c|}{\textbf{Value}} & \multicolumn{1}{c|}{\textbf{\begin{tabular}[c]{@{}c@{}}Memory\\ Usage\\ (MB)\end{tabular}}} & \multicolumn{1}{c|}{\textbf{\begin{tabular}[c]{@{}c@{}}CPU\\ Usage\end{tabular}}} \\ \hline \hline
\textbf{No Virtualization} & {\color[HTML]{CB0000} \textbf{1771.58}} & {\color[HTML]{036400} 2841} & {\color[HTML]{CE6301} 0.63\%} & {\color[HTML]{CB0000} \textbf{33714.08}} & {\color[HTML]{036400} 2839} & {\color[HTML]{CE6301} 0.63\%} & {\color[HTML]{CB0000} \textbf{35408.98}} & {\color[HTML]{036400} 2837} & {\color[HTML]{CE6301} 0.63\%} & {\color[HTML]{CB0000} \textbf{2812135.26}} & {\color[HTML]{036400} 2838} & {\color[HTML]{CE6301} 0.63\%} & {\color[HTML]{CB0000} \textbf{1135.37}} & {\color[HTML]{036400} 2842} & {\color[HTML]{CE6301} 0.63\%} \\ \hline
\textbf{Docker} & {\color[HTML]{CB0000} \textbf{2071.79}} & {\color[HTML]{036400} 2791} & {\color[HTML]{CE6301} 8.79\%} & {\color[HTML]{CB0000} \textbf{32232.27}} & {\color[HTML]{036400} 2792} & {\color[HTML]{CE6301} 8.76\%} & {\color[HTML]{CB0000} \textbf{51522.74}} & {\color[HTML]{036400} 2794} & {\color[HTML]{CE6301} 8.76\%} & {\color[HTML]{CB0000} \textbf{2329719.99}} & {\color[HTML]{036400} 2794} & {\color[HTML]{CE6301} 8.76\%} & {\color[HTML]{CB0000} \textbf{1581.1}} & {\color[HTML]{036400} 2796} & {\color[HTML]{CE6301} 8.76\%} \\ \hline
\textbf{VMware} & {\color[HTML]{CB0000} \textbf{1180.92}} & {\color[HTML]{036400} 1337} & {\color[HTML]{CE6301} 3.52\%} & {\color[HTML]{CB0000} \textbf{12860.3}} & {\color[HTML]{036400} 1330} & {\color[HTML]{CE6301} 3.52\%} & {\color[HTML]{CB0000} \textbf{13167.09}} & {\color[HTML]{036400} 1336} & {\color[HTML]{CE6301} 3.52\%} & {\color[HTML]{CB0000} \textbf{568262.62}} & {\color[HTML]{036400} 1338} & {\color[HTML]{CE6301} 3.52\%} & {\color[HTML]{CB0000} \textbf{482.88}} & {\color[HTML]{036400} 1340} & {\color[HTML]{CE6301} 3.52\%} \\ \hline
\textbf{KVM} & {\color[HTML]{CB0000} \textbf{1406.08}} & {\color[HTML]{036400} 431} & {\color[HTML]{CE6301} 4.50\%} & {\color[HTML]{CB0000} \textbf{12969.73}} & {\color[HTML]{036400} 434} & {\color[HTML]{CE6301} 4.50\%} & {\color[HTML]{CB0000} \textbf{13192.78}} & {\color[HTML]{036400} 440} & {\color[HTML]{CE6301} 4.50\%} & {\color[HTML]{CB0000} \textbf{908660.05}} & {\color[HTML]{036400} 440} & {\color[HTML]{CE6301} 4.50\%} & {\color[HTML]{CB0000} \textbf{491.73}} & {\color[HTML]{036400} 448} & {\color[HTML]{CE6301} 4.50\%} \\ \hline
\end{tabular}
\end{table*}

\begin{figure}[h]
\captionsetup[subfloat]{farskip=2pt,captionskip=1pt}
	\centering
   
    \subfloat[]{
		\label{fig:Blogbench_read}
		\includegraphics[width=.22\textwidth]{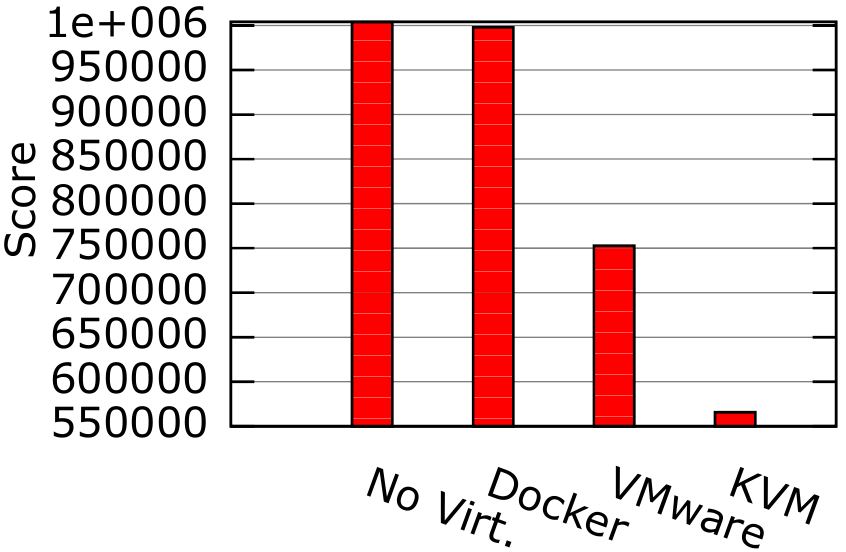}%
	}		
	\hspace{0.0cm}
	\subfloat[]{
		\label{fig:Blogbench_write}
		\includegraphics[width=.23\textwidth]{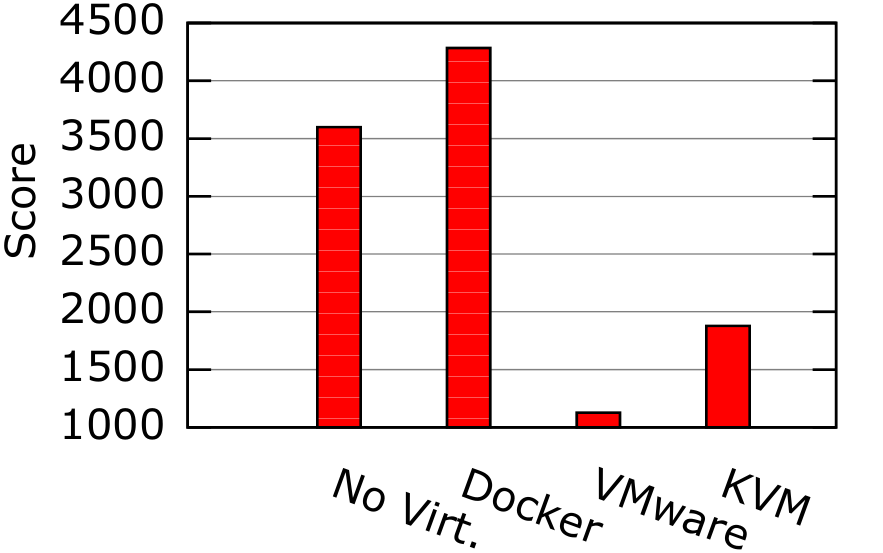}	}

		\caption{(a) Blogbench test read results. (b) Blogbench test write results.}
 	\label{fig:Blogbench} 
\end{figure}

\begin{figure}[h]
     \centering
      \includegraphics[width=.6\linewidth]{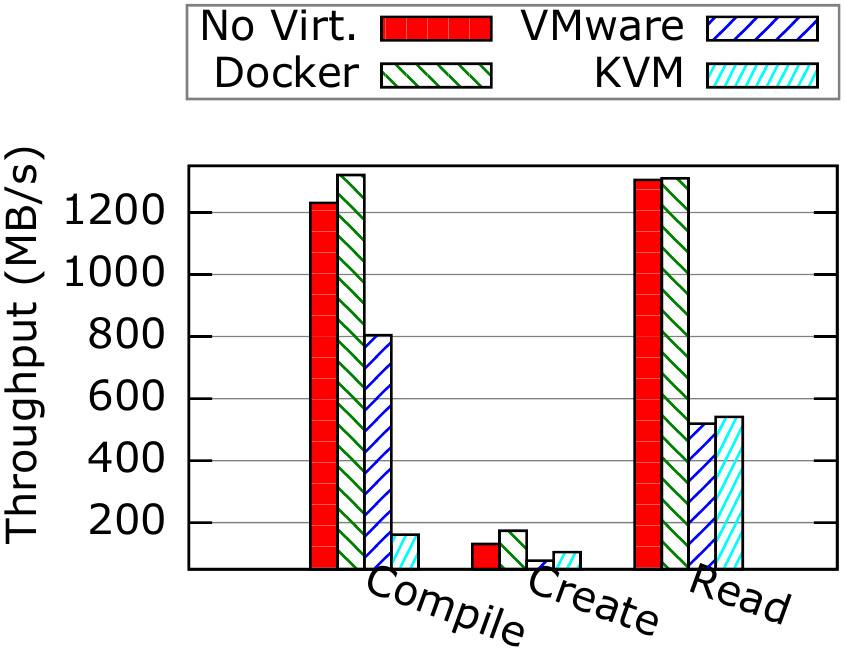}
      \caption{Compilebench results. "Compile" indicates the Compile test. "Create" refers to the Initial Create test. "Read" specifies the Read Compiled Tree test}
      \label{fig:compileBench}
\end{figure}

To assess how the four testing platform would perform against real-world server operations, we use the \textbf{pts/blogbench} benchmark. The results are recorded separately for read and write operations (see Figure \ref{fig:Blogbench_read}). A higher "Score" is equivalent to better performance.
In the previous experiment, Docker had the best disk access throughput. Consequently, the initial assumption was to expect Docker to come on top again. However, this time it obtained pretty much the same score as the native execution, both far above VMWare and KVM. This trend is repeated when executing the \textbf{pts/compilebench} benchmark (see Figure \ref{fig:compileBench}). On the \textit{compile} phase, Docker has a throughput of 1320 MB/s while the native execution achieves 1230 MB/s. This nevertheless does not necessarily means that Docker runs faster than the native as the recorded values are averaged by the benchmark suite. When studying the standard deviation returned by each runs, the native shows a deviation of 40.71 MB/s and Docker comes with 22.89 MB/s, which put the two execution in a fairly close range. On the same test, VMware obtained an average throughput that is 804.74 MB/s and KVM achieved 161.3 MB/s.
Overall, these IO stressing benchmarks show that the native execution and Docker perform quite similarly. These results were expected as Docker directly runs on the host. However, the experiments also showed that IO-intensive applications hurt VMs. It is explained by the fact that VMMs consume significant run-time resources to handle context switches as the benchmarks continually attempt accessing the hardware through system calls. Overall, managing IO device accesses degrades the performance of the applications running in the VMs.

\subsubsection{Processor Performance}
In this section, we discuss experiments related to comparing processor performance. The main goal is to assess how fast the compute-intensive applications will run under the four test environments (native, Docker, VMWare, and KVM). We start with the \textbf{pts/aobench} benchmark.
\begin{figure}[h]
     \centering
      \includegraphics[width=.6\linewidth]{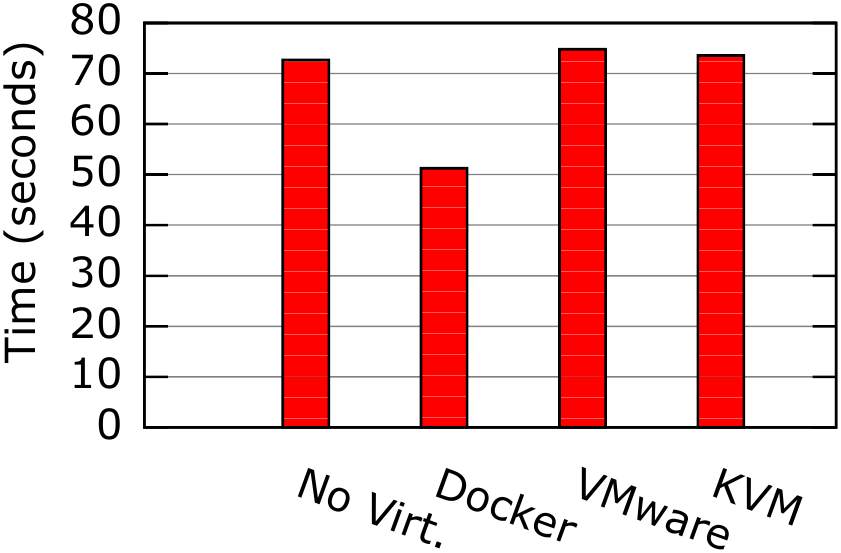}
      \caption{aobench execution results}
      \label{fig:aobench}
\end{figure}
The results are presented in Figure \ref{fig:aobench}. Here again, Docker achieves the best performance as the benchmark execution completes within 51 seconds. The native execution, VMWare, and KVM respectively terminate after 72 seconds, 74 seconds, and 73 seconds. The experiment also reveals that VMs perform like native execution. The similar performance observed between the native execution, VMware and KVM are expected as compute-intensive workloads typically use less privileged instructions than IO-intensive applications. VMs running on modern VMMs directly execute on the underlying CPU. The VMM is only invoked when a VM issues privileged instructions.

When we run the \textbf{pts/asmfish} benchmark the native execution comes at the top followed by Docker.
\begin{figure}[h]
     \centering
      \includegraphics[width=.6\linewidth]{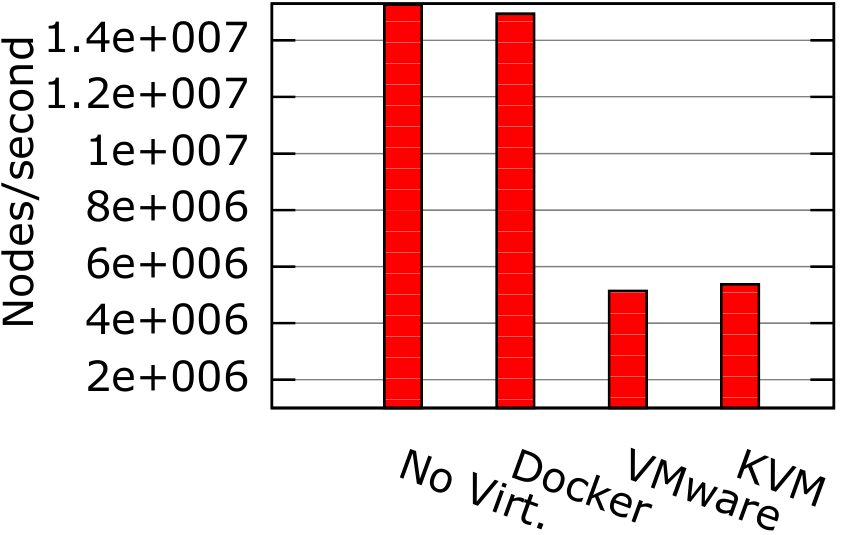}
      \caption{asmfish execution results}
      \label{fig:asmfishBench}
\end{figure}
As opposed to the results observed with \textbf{pts/aobench}, the VMware and KVM performances are quite similar but far lower than that of Docker and the native execution. After investigation, it appears that the performance of \textbf{pts/aobench} depends on the available processing cores \cite{asmfish}. The benchmark description also highlights the fact that there is a significant performance improvement when increasing the number of AMD processors instead of Intel processors. In our testing setup, the native execution and Docker ran with 16 AMD cores while KVM and VMware only used only four cores, justifying the performance difference.

Overall, the lesson here is similar to what was observed with IO-intensive benchmarks. The native execution and Docker generally come at the top, and KVM and VMware have lower performance, but remain close in execution profile.

\subsubsection{System Performance}
Finally, we discuss experiments related to comparing system-level performances. We start by running \textbf{pts/phpbench}. It runs tests that assess several aspects of the workloads typically observed in PHP servers.
\begin{figure}[h]
     \centering
      \includegraphics[width=.6\linewidth]{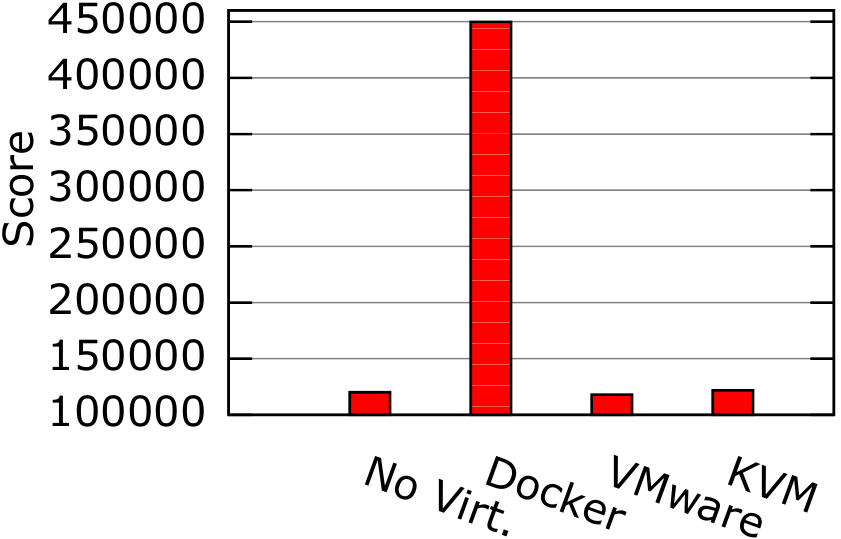}
      \caption{phpbench execution results}
      \label{fig:phpbench}
\end{figure}
Figure \ref{fig:phpbench} summarizes the finding. The native execution, VMWare and KVM have similar score while Docker performs way better. Table \ref{tab:stress_ng} summarizes execution results from running \textbf{pts/stress-ng}. Overall the results show that once more the native execution and Docker have similar performances that are above that of VMWare and KVM. Nevertheless, KVM and VMware seems to be less memory hungry.

As a summary, throughout the experiments that we carried out, we observed that Docker and the native execution performed better on compute-intensive, IO-intensive, and system benchmarks. In general, VMware and KVM had lower performance regardless of the stressor we used. The fact is VMs are more suited for deploying systems as they can run OSes, nested VMs, and even containers. However, they incur significant performance degradation compared to containers and bare-metal systems. Containers showed near-native performance as they are similar to the processes that execute directly above the host OS but are mostly limited to running specific applications.

\section{Conclusion}
We studied three types of virtualization environments in this work: containers, Type-1, and Type-2 virtual machine monitors. We evaluated the performance achieved by applications running in Docker, KVM, and VMWare Workstation. One of the main observations is that containers appear to be the best platform to run applications if the target is fast execution with low overhead. However, they are not suited for deploying complete systems as they can only run applications. We also observed that VMware and KVM tend to have similar execution performances with minor differences. Future work will extend the study to other virtualization systems such as Xen and explore the performance achieved when virtualization technologies are nested. 

\bibliographystyle{ACM-Reference-Format}
\bibliography{acmart}

\end{document}